\documentclass[cits]{PoS} 
\title{Instantaneous Bethe--Salpeter Approach to Pseudoscalar
Mesons}\ShortTitle{Instantaneous Bethe--Salpeter Approach to
Pseudoscalar Mesons}\author{\speaker{Wolfgang Lucha}\\Institute
for High Energy Physics, Austrian Academy of Sciences,
Nikolsdorfergasse 18, A-1050 Vienna, Austria\\E-mail:
\email{Wolfgang.Lucha@oeaw.ac.at}}

\abstract{Light pseudoscalar mesons are Janus-type particles:
Within quantum chromodynamics, they must be described as bound
states of its fundamental degrees of freedom and as the (pseudo-)
Goldstone bosons of its spontaneously broken chiral symmetry. This
janiform nature of pions and kaons may be easily accommodated by
the Bethe--Salpeter formalism in its instantaneous limit: Starting
from the general shape of the Bethe--Salpeter solutions for light
pseudoscalar mesons at large Euclidean momenta, we provide the
exact relationship between the solutions of our bound-state
equation and the underlying interactions, boiled down to
potentials $V(r)$ depending on the interquark distance $r.$ For
massless quarks, $V(r)$ exhibits, at the origin, a
(logarithmically softened) Coulomb singularity crucial for
counterbalancing all positive contributions to the bound-state
mass but rises, for large $r,$ to infinity and can hence be
regarded as confining. For massive quarks, $V(r)$ still features a
similar (logarithmically softened) Coulomb singularity at the
origin; for quark masses too large, however, the potential's
confining character gets lost: $V(r)$ approaches, for large $r,$ a
nonpositive~finite limit.

\begin{center}\begin{tabular}{cc}
\includegraphics[scale=.751716]{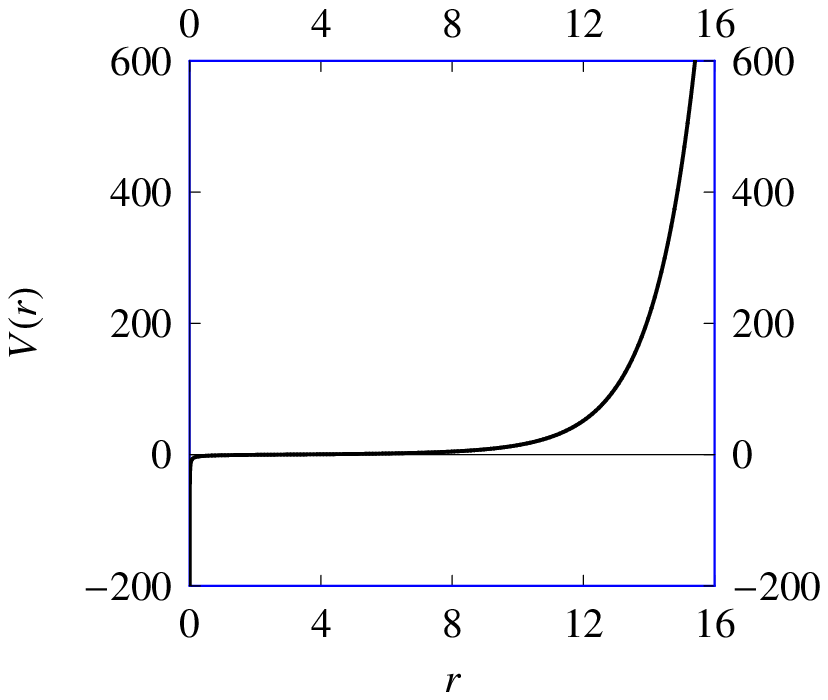}&
\includegraphics[scale=.751716]{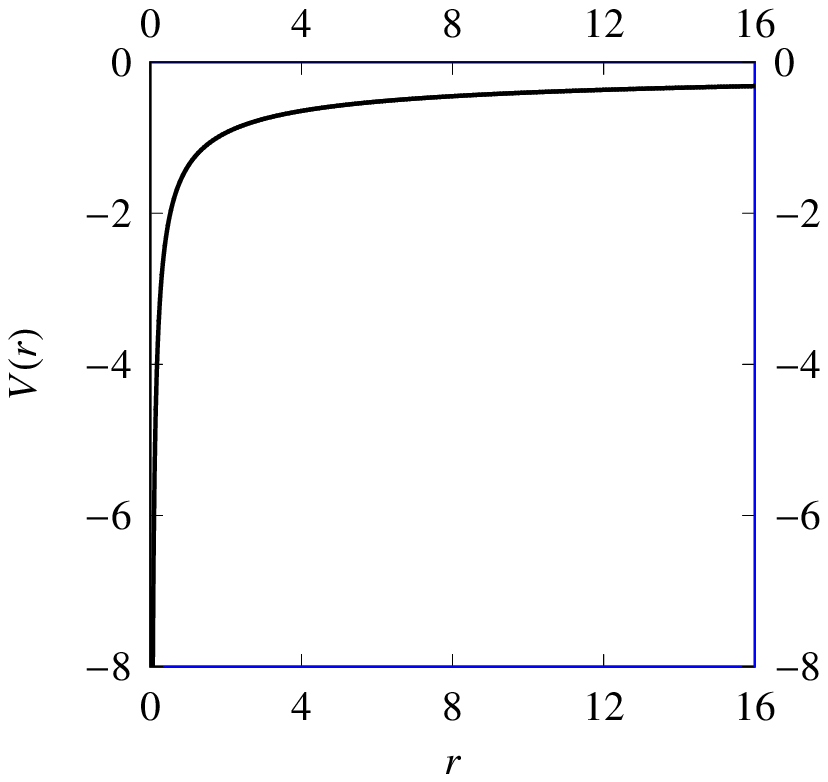}
\end{tabular}\end{center}}

\FullConference{The European Physical Society Conference on High
Energy Physics -- EPS-HEP2013\\18--24 July 2013\\Stockholm,
Sweden}

\begin{document}\section*{Introduction, Cursory Sketch of Basic
Idea, Summary of Findings, and Conclusions}In principle, the
homogeneous {\em Bethe--Salpeter equation\/} \cite{BSE} provides a
quantum-field-theoretic description of relativistic bound states
in Minkowski space. In real life, practical or even conceptual
obstacles often prompt us to content ourselves with {\em
three-dimensional reductions\/} of this formalism. Most prominent
among the resulting bound-state equations is the {\em Salpeter
equation\/} \cite{SE}, obtained by assuming that all bound-state
constituents interact instantaneously and propagate freely,
cf.~Ref.~\cite{NR}. An astounding drawback of the Salpeter
equation is that, depending on the nature of the interactions
involved, it predicts, even in situations where we expect to
obtain just stable bound states, also states that develop
instabilities. This issue has been thoroughly analysed for the
Salpeter equation without negative-energy contributions and a
generalization \cite{IBSE} thereof \cite{R(IB)SE-Stab}, and the
full Salpeter equation \cite{SE-Stab}. As a by-product, such
studies bore out the need for {\em exact analytic solutions\/} of
the~Salpeter~equation.

Examples of these may be found by inverting the procedure:
Assuming spherical symmetry, all interactions are encoded in
configuration-space central potentials and Salpeter's equation
reduces to systems of radial eigenvalue equations \cite{RE}. Then
rigorous relations between the properties of bound states and the
interactions which their constituents experience can by
established by determining {\em for pre-selected solutions\/}
those potentials for which the bound-state equation yields these
solutions~\cite{ES}.

Naturally, the question arises how that relation looks like for
{\em light pseudoscalar mesons\/} viewed as --- due to (explicitly
and) spontaneously broken global symmetries of quantum
chromodynamics (almost) massless --- quark--antiquark bound
states. In the {\em chiral limit\/} of quantum chromodynamics with
{\em dynamically broken\/} chiral symmetry, pseudoscalar-meson
Bethe--Salpeter solutions \cite{MRT} fall off (for large Euclidean
momenta) like the inverse fourth power of the quarks' relative
four-momentum. Choosing our Salpeter model according to these
insights and Fierz-symmetric effective interactions of the quarks
then gives the central potentials for massless and massive quarks
in analytic form \cite{PM}.

\end{document}